\documentstyle[twocolumn,aps,epsf,floats]{revtex}

\begin{document}

\newcommand{\bi}[1]{\mbox{\boldmath$#1$}}

\draft

\title{Asymptotic power law of moments in a random multiplicative
  process \\ with weak additive noise}

\author{Hiroya Nakao}

\address{Department of Physics, Graduate School of Sciences,
  Kyoto University, Kyoto 606, Japan}

\date{January 31, 1998}

\maketitle

\begin{abstract}
  It is well known that a random multiplicative process with weak
  additive noise generates a power-law probability distribution.
  It has recently been recognized that this process exhibits another
  type of power law: the moment of the stochastic variable scales as a
  function of the additive noise strength.
  We clarify the mechanism for this power-law behavior of moments by
  treating a simple Langevin-type model both approximately and
  exactly, and argue this mechanism is universal.
  We also discuss the relevance of our findings to noisy on-off
  intermittency and to singular spatio-temporal chaos recently
  observed in systems of non-locally coupled elements.
\end{abstract}


\section{Introduction}

Power laws are observed in a wide variety of natural phenomena and
mathematical models. Some examples are the critical behavior near
second order phase transitions, Kolmogorov's law of fully developed
turbulence, size distribution of avalanches in models of
self-organized criticality, Gutenberg-Richter's law of earthquakes,
distribution of price fluctuations in economic activities, and Zipf's
law in linguistics. Clarifying the mechanisms for the emergence of
these power laws has long been a subject of many challenges.

The random multiplicative process (RMP) is a well-known mechanism
leading to power-law behavior. It is a stochastic process where the
stochastic variable is driven by a multiplicative noise. It has been
extensively used as a model for a variety of systems such as on-off
intermittency~\cite{Fujisaka,Yamada,Pikovsky,Platt,Venkataramani,Cenys},
lasers~\cite{Graham}, economic activity~\cite{Levy,Takayasu},
variation of biological populations in fluctuating
environment~\cite{Turelli}, and passive scalar field advected by
fluid~\cite{Deutsch}.

In real systems, the stochastic variable may often be driven not only
by the multiplicative noise, but also by some weak additive noise.
This weak additive noise becomes important when the amplitude of the
stochastic variable $x$ takes small values, and introduces an
effective lower bound of $x$. Actually, this lower bound may be
crucial, because it guarantees the existence of a stationary
probability distribution function (PDF).
Furthermore, the PDF here has a power-law form over a wide range of
$x$~\cite{Fujisaka,Pikovsky,Venkataramani,Levy,Takayasu,Deutsch,Kuramoto2}.

For example, Venkataramani {\it et al.}~\cite{Venkataramani}
introduced a Langevin equation with multiplicative and additive noise
terms as a model for noisy on-off intermittency. They obtained a
stationary PDF with a power-law tail.
The same form of Langevin equation was treated by Takayasu
{\it et al.}~\cite{Takayasu} as a model for economic activity, and they also
showed that the PDF obeys a power law.
A similar model was introduced by Levy {\it et al.}~\cite{Levy}:
it describes a discrete stochastic process driven by a
multiplicative noise. They introduced a lower bound to the stochastic
variable explicitly, and showed again that the PDF obeys a power law.
Venkataramani {\it et al.} and Takayasu {\it et al.} treated the
additive noise explicitly, while the lower bound introduced by Levy
{\it et al.} plays a role similar to the additive noise.
In this paper, we are concerned with this type of stochastic processes.

Recently, another type of asymptotic power law was found in the
above type of stochastic processes. In previous
papers~\cite{Kuramoto1,Kuramoto2}, we introduced a stochastic process
in order to explain the power law displayed by the spatial correlation
function $C(r)$, i.e., $C(r) \simeq C_0 - C_1 r^{\alpha}$ for small
enough $r$, observed in the spatio-temporal chaotic regime of systems
with non-locally coupled elements.
Our explanation was based on a RMP with weak additive noise such as
described above. Note, however, that the power-law correlation here is
not a direct result of the power-law tail of the PDF itself, but it is
a result of the asymptotic power law of moments $<x^q>$ of the stochastic
variable $x$ as a function of the strength $s$ of the additive noise,
i.e., $ <x^q> \simeq G_0 + G_1 s^{H(q)}$.
This gives another mechanism leading to power-law behavior in such
stochastic processes.

The goal of this paper is to clarify this mechanism for the emergence
of the power law of moments with respect to the strength of the
additive noise. We achieve this by using a simple Langevin-type model,
and argue that the mechanism proposed is a universal one in generating
various power laws.

The outline of this paper is as follows: In Sec. II we introduce the
model to be studied, and display its typical behavior.
In Sec. III we treat the model approximately in order to outline the
mechanism for the emergence of the power law of moments,
and then exactly in Sec. IV.
In Sec. V we discuss the robustness of the power law with regard to
boundary conditions and nature of the noise.
We also show some results obtained by numerical calculations
with colored noises.
In Sec. VI we discuss an application of our theory to noisy on-off
intermittency. As an example, we show a result obtained by a numerical
calculation of coupled chaotic elements. Furthermore, we discuss the
relation of the power law of moments to the power-law spatial
correlations observed in systems of non-locally coupled chaotic elements.
We summarize our results in Sec. VII.

\section{Analytical model}

\subsection{Langevin equation}

As a model for a RMP with weak additive noise, we employ a Langevin
equation
\begin{equation}
  \label{eq:Langevin}
  \frac{dx(t)}{dt} = \lambda(t) x(t) + \eta(t) ,
\end{equation}
where $x(t)$ is a stochastic variable, $\lambda(t)$ a multiplicative
noise, and $\eta(t)$ an additive noise.
We assume both $\lambda(t)$ and $\eta(t)$ to be Gaussian-white, and
their average and variance to be given by
\begin{equation}
  \begin{array}{c}
    <\lambda(t)> = \lambda_0, \ 
    < \left[ \lambda(t) - \lambda_0 \right] \left[ \lambda(t') - \lambda_0 \right] >
    = 2 D_{\lambda} \delta(t-t') , \\
    \\
    <\eta(t)> = 0, \ <\eta(t) \ \eta(t')> = 2 D_{\eta} \delta(t-t') .
  \end{array}
\end{equation}
We further assume $D_{\eta} \ll D_{\lambda}$, namely, the additive noise
is sufficiently weaker than the multiplicative noise.

This simple Langevin equation (\ref{eq:Langevin}) has been widely
used in many studies of various
systems~\cite{Venkataramani,Graham,Takayasu,Deutsch}.
The physical meaning of $x(t)$, $\lambda(t)$, and $\eta(t)$ may be
different depending on the specific system under consideration. For
example, in the case of lasers, $x(t)$ gives the number of photons,
$\lambda(t)$ its fluctuating amplification rate, and $\eta(t)$ the
noise due to random spontaneous emissions of atoms.
When we are working with noisy on-off intermittency, $x(t)$ gives the
measure of a distance from the invariant manifold, $\lambda(t)$ the
instantaneous vertical Lyapunov exponent, and $\eta(t)$ the noise due
to a parameter mismatch or some other cause. In the context of
economic models, $x(t)$ represents the wealth, $\lambda(t)$ the rate
of change of the wealth, and $\eta(t)$ some external noise of various
sources.

\subsection{Boundary conditions}

In order to obtain a statistically stationary state from the Langevin
equation (\ref{eq:Langevin}), we generally need upper and lower bounds
of $x$.
In our model, the weak additive noise may act as an effective lower
bound.

\paragraph*{(a) Lower bound}

Without the additive noise, $x(t)$ tends to $0$ when the average
expansion rate $\lambda_0$ is negative. The additive noise introduces
an effective lower bound of $x(t)$, which keeps $x(t)$ away from the
zero value even if $\lambda_0 < 0$.
In this paper, we treat the additive noise explicitly, while in some
other studies it is replaced by a reflective wall (infinitely high
barrier) placed at some small $x$.

\paragraph*{(b) Upper bound}

To be realistic, when $x(t)$ takes too large values, it should be
saturated by some effect such as nonlinearity. We simply introduce
this effect as boundary conditions, specifically reflective walls at
$x = \pm 1$.

\begin{figure}[htbp]
  \begin{center}
    \leavevmode
    \epsfxsize=8cm
    \epsffile{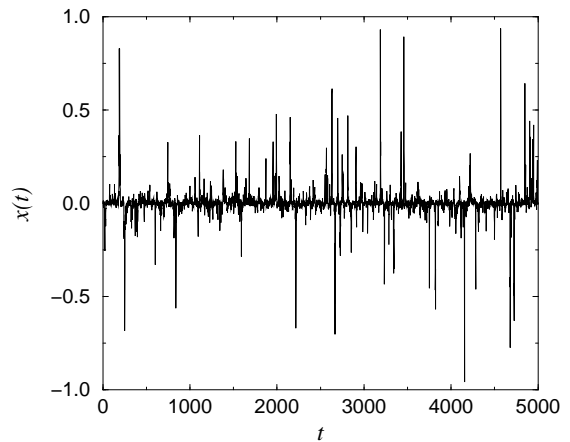}
  \end{center}
  \caption{Typical time evolution of the amplitude $x(t)$
    of the Langevin equation (1) for $\lambda_0 = -0.5$,
    $D_\lambda = 0.5$, $D_\eta = 0.00005$.}
  \label{fig:1}
\end{figure}

With these upper and lower bounds provided by an additive noise and
reflective walls, the Langevin equation (\ref{eq:Langevin}) admits a
statistically stationary state.
In Fig.~\ref{fig:1}, we show a typical time evolution of $x(t)$
governed by the Langevin equation (\ref{eq:Langevin}) for slightly
negative $\lambda_0$, where the reflective walls are placed at $x =
\pm 1$.
In spite of the negative average expansion rate $\lambda_0$, $x(t)$
does not simply decay but exhibits intermittent bursts. The generation
of bursts may be interpreted as follows.
Due to the weak additive noise, $x(t)$ may generally have small but
finite values. If positive $\lambda(t)$ happens to persist over some
period, $x(t)$ will be amplified exponentially and attain large
values, which is nothing but bursts. Of course, $x(t)$ may eventually
decay to the noise level because of negative $\lambda_0$.
The chance of bursts will increase with the additive noise strength.
Why this leads to a power-law dependence of moments of $x(t)$ on the
additive noise strength can be understood from the argument below.

The intermittency described above has the same statistical nature as
the noisy on-off intermittency. Actually, one may consider the noisy
on-off intermittency as a stochastic process of the type described
above.

\newpage

\section{Approximate treatment}

In order to give an outline of the mechanism underlying the emergence
of the power law of moments, we first treat the Langevin equation
(\ref{eq:Langevin}) approximately.

\subsection{Fokker-Planck equation}

We introduce a characteristic amplitude
\begin{equation}
  s = \sqrt{\frac{D_\eta}{D_\lambda}} \ \ \ (0 < s < 1)
\end{equation}
which results from the balance between the fluctuation due to the
multiplicative noise \mbox{$<(\lambda x)^2>$}$ \sim D_{\lambda} x^2$ and
the fluctuation due to the additive noise \mbox{$<\eta^2>$}$ \sim D_{\eta}$.
We divide the range of $x$ into two parts, $0 < |x| < s$ and $s < |x|
< 1$, and ignore in each region one of the noise sources which is less
dominant there.
Since the system is statistically symmetric with respect to the
transformation $x \rightarrow -x$, we consider only the absolute value
$|x|$ hereafter.

\paragraph*{(a) $s < |x| < 1$}

In this region, we ignore the effect of the additive noise and
consider a Langevin equation
\begin{equation}
  \label{eq:Langevin2}
  \frac{dx(t)}{dt} = \lambda(t) x(t) .
\end{equation}
By introducing a new variable $y(t) = \log |x(t)|$,
Eq. (\ref{eq:Langevin2}) is rewritten as
\begin{equation}
  \frac{dy(t)}{dt} = \lambda(t) ,
\end{equation}
which gives a diffusion process with mean drift.

Let $P_1(x, t)$ denote the PDF of $x(t)$ and ${\bar P_1}(y, t)$ the
corresponding PDF of $y(t)$. The Fokker-Planck equation corresponding
to Eq. (\ref{eq:Langevin2}) takes the form
\begin{equation}
  \frac{\partial}{\partial t} {\bar P_1}(y, t) 
  = -\frac{\partial}{\partial y} j_1(y, t) ,
\end{equation}
where the flux $j_1(y, t)$ is given by
\begin{equation}
  j_1(y, t) = \lambda_0 {\bar P_1}(y, t) - D_{\lambda} \frac{\partial}{\partial y} 
  {\bar P_1}(y, t) .
\end{equation}
Setting reflective walls at $x = \pm 1$ is equivalent to assuming a
no-flux boundary condition for $j_1$ at $y = 0$, i.e., $j_1(y = 0, t)
= 0$.

\paragraph*{(b) $0 < |x| < s$}

In this region, we ignore the effect of the multiplicative noise, and
this leads to a Langevin equation
\begin{equation}
  \label{eq:Langevin3}
  \frac{dx(t)}{dt} = \eta(t) .
\end{equation}

Let $P_2(x, t)$ be the PDF of $x(t)$. $P_2(x, t)$ obeys the
Fokker-Planck equation
\begin{equation}
  \frac{\partial}{\partial t} P_2(x, t) 
  = - \frac{\partial}{\partial x} j_2(x, t) ,
\end{equation}
where the flux $j_2(x, t)$ is now defined by
\begin{equation}
  j_2(x, t) = - D_{\eta} \frac{\partial}{\partial x} P_2(x, t) .
\end{equation}
The boundary conditions to be imposed here are the continuity of $P_1$
and $P_2$, and also $j_1$ and $j_2$, each at $|x| = s \ ( y = \log s )$.

\subsection{Stationary PDF with a power-law tail}

We calculate here the stationary solution of the Fokker-Planck
equation.

\paragraph*{(a) $s < |x| < 1$}

Stationarity condition ${\partial {\bar P_1}}/{\partial t} = 0$ gives
${\partial j_1}/{\partial y} = 0 $, i.e., $j_1(y) \equiv const.$, and the
no-flux boundary condition $j_1(y = 0) = 0$ gives $j_1(y) \equiv 0$.
Therefore, the stationary solution ${\bar P_1}(y)$ satisfies
\begin{equation}
  0 = \lambda_0 {\bar P_1}(y) - D_{\lambda} \frac{\partial}{\partial y} 
  {\bar P_1}(y) .
\end{equation}
This can be solved as
\begin{equation}
  {\bar P_1}(y) = C \exp\left(\frac{\lambda_0}{D_{\lambda}} y \right) ,
\end{equation}
where $C$ is a normalization constant.
In terms of the original variable $x$, we obtain
\begin{eqnarray}
  P_1(x) &=& {\bar P_1}(y) \frac{dy}{dx} \cr
  &=& C \frac{1}{|x|} \exp\left(\frac{\lambda_0}{D_{\lambda}} \log(|x|) \right) \cr
  &=& C |x|^{\frac{\lambda_0}{D_{\lambda}}-1} .
\end{eqnarray}
Thus, the PDF obeys a power law in this region. The exponent of this
power law is determined by the ratio of $\lambda_0$ to $D_{\lambda}$,
i.e., by the basic statistical characteristics of the multiplier
$\lambda(t)$, and does not depend on the nature of the additive noise.
We denote this ratio as $\beta$ hereafter:
\begin{equation}
  \beta = \frac{\lambda_0}{D_{\lambda}} .
\end{equation}

\paragraph*{(b) $0 < |x| < s$}

A general form of the stationary solution is given by $ P_2(x) = A x +
B $, where $A$ and $B$ are constants.
Continuity of the flux and the PDF at $|x| = s$, i.e., $j_2(s) =
j_1(s) \equiv 0$ and $P_2(s) = P_1(s)$ gives $A=0$ and $B = P_1(s)$.
Therefore, $P_2(x)$ takes a constant value:
\begin{equation}
  P_2(x) \equiv P_1(s) = C s^{\frac{\lambda_0}{D_{\lambda}}-1} .
\end{equation}

\begin{figure}[htbp]
  \begin{center}
    \leavevmode
    \epsfxsize=8cm
    \epsffile{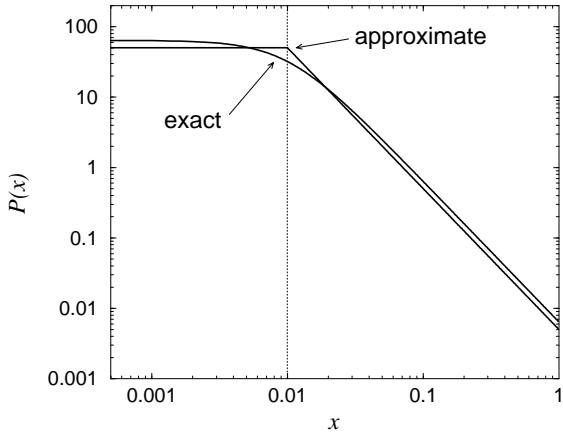}
  \end{center}
  \caption{PDFs $P(x)$ vs. $x$, where the approximate and exact curves are compared.
    The parameters are the same as in Fig. 1.}
  \label{fig:2}
\end{figure}

Finally, the approximate stationary PDF is obtained as
\begin{equation}
  \label{eq:PDF1}
  P(x) = \left\{
    \begin{array}{cc}
      C s^{\beta-1} & ( 0 < |x| < s ) \\
      \\
      C x^{\beta-1} & ( s < |x| < 1 ) \\
      \\
      0 & ( |x| > 1 )
    \end{array}
  \right. .
\end{equation}
The normalization constant $C$ is determined from
\begin{equation}
  \int_{-1}^{1} P(x) dx = 2 \int_{0}^{1} P(x) dx = 1 ,
\end{equation}
and calculated as
\begin{eqnarray}
  C & = & \left[ 2 \left( \int_{0}^{s} s^{\beta-1} dx + \int_{s}^{1}
      x^{\beta-1} dx \right) \right]^{-1} \cr
  & = & \left[ 2 \left(
      s^{\beta} + \frac{1-s^{\beta}}{\beta} \right) \right]^{-1} .
\end{eqnarray}
Thus, the PDF consists of three parts, i.e., a constant part near
the origin where $x(t)$ is dominated by a normal diffusion process,
a power-law tail where $x(t)$ is dominated by a RMP, and a
vanishing part.
The boundary between the constant part and the power-law tail is
located at $|x| = s$, which is proportional to the additive noise
strength $\sqrt{D_{\eta}}$.
We show this approximate PDF (\ref{eq:PDF1}) in Fig.~\ref{fig:2}.

\subsection{Moments}

The $q$-th moment \mbox{$<x^q>$} of $|x|$ in this stationary state
(\ref{eq:PDF1}) is calculated as
\begin{eqnarray}
  <x^q> & = & \int_{-1}^{1} |x|^q P(x) dx = 2 \int_{0}^{1} x^q P(x) dx \cr
  & = & 2 C \left( \int_{0}^{s} x^q s^{\beta-1} dx
    + \int_{s}^{1} x^q x^{\beta-1} dx \right) \cr
  & = & \frac{\beta}{\beta+q}
  \frac{1 + (\frac{\beta+q}{1+q}-1) s^{\beta+q}}{1 + (\beta-1) s^\beta} .
\end{eqnarray}
We can write the above in the form
\begin{equation}
  \label{eq:Mom_form}
  <x^q> = c_q \frac{1 + a_q s^{\beta+q}}{1 + a_0 s^{\beta}} ,
\end{equation}
where $c_q$ and $a_q$ are given by
\begin{equation}
  c_q = \frac{\beta}{\beta+q},\ \ \  a_q = \frac{\beta+q}{1+q} - 1.
\end{equation}
Note that the exponent $\beta$ of the PDF now appears as the power of
$s$. As we explain below, the form of Eq. (\ref{eq:Mom_form}) is all
we need for the emergence of the asymptotic power law of moments.

\subsection{Asymptotic forms of the moments}

We investigate the asymptotic forms of the moment \mbox{$<x^q>$} in
the limit of small additive noise $s$. We consider only the
practically interesting case of positive $q$.

\paragraph*{(a) $\beta > 0$}

By expanding the denominator of \mbox{$<x^q>$} and taking the lowest order in
$s$, we obtain
\begin{equation}
  \label{eq:Mom_asym1}
  <x^q> \simeq c_q \left( 1 - a_0 s^{\beta} \right) .
\end{equation}

\paragraph*{(b) $\beta < 0$}

Ignoring $1$ in the denominator of \mbox{$<x^q>$}, we obtain
\begin{equation}
  \label{eq:Mom_asym2}
  <x^q> \simeq \frac{c_q}{a_0} \left( s^{|\beta|} + a_q s^q \right) .
\end{equation}
Which of $|\beta|$ and $q$ is smaller determines which of the two
terms on the right-hand side of Eq. (\ref{eq:Mom_asym2}) dominates.
We thus obtain
\begin{equation}
  \label{eq:Mom_asym3}
  <x^q> \simeq \left\{
    \begin{array}{ll}
      \displaystyle{ \frac{c_q}{a_0} s^{|\beta|} } & (|\beta| < q) \\
      \\
      \displaystyle{ \frac{c_q a_q}{a_0} s^q } & (|\beta| > q)
    \end{array}
  \right. .
\end{equation}

These results show that the moment \mbox{$<x^q>$} approaches a simple
power-law form as a function of the position $s$ of the boundary, or
the strength $\sqrt{D_{\eta}}$ of the additive noise:
\begin{equation}
  \label{eq:Power}
  <x^q> \simeq G_0 + G_1 s^{H(q)} ,
\end{equation}
where $G_0$ and $G_1$ are constants.
$G_0$ vanishes when $\beta < 0$ ($\lambda_0 < 0$), while it takes a
finite value when $\beta > 0$ ($\lambda_0 > 0$). Thus, we have
obtained the power law of moments.

\subsection{Exponents}

The exponent $H(q)$ of the moment \mbox{$<x^q>$} is determined by $\beta$,
namely, the ratio of $\lambda_0$ to $D_{\lambda}$. From
Eqs. (\ref{eq:Mom_asym1}) and (\ref{eq:Mom_asym3}), $H(q)$ varies with
$q$ as follows:

{\it (a) $\beta < 0$}

\begin{equation}
  \label{eq:Exp_neg}
  H(q) = \left\{
    \begin{array}{cc}
      q & (0 < q < |\beta|) \\
      \\
      |\beta| & (|\beta| < q)
    \end{array}
  \right. ,
\end{equation}

{\it (b) $\beta > 0$}

\begin{equation}
  \label{eq:Exp_pos}
  H(q) = \beta .
\end{equation}
We notice that $H(q) = q$ when $0 < q < |\beta|$, but $H(q) = |\beta|$
when $|\beta| < q$ or $\beta > 0$ without dependence on $q$.

\subsection{Other asymptotic regimes}

When $\beta \simeq 0$ or $|\beta| \simeq q$, there exist other
asymptotic regimes where the asymptotic form of the moment in the $s
\rightarrow 0$ limit is not a power law.

\paragraph*{(a) $\beta \simeq 0$}

Consider a parameter region where $\beta \simeq 0$ and $| \beta \log s
| \ll 1 $. The denominator of \mbox{$<x^q>$} can be expanded as
\begin{eqnarray}
  1 + a_0 s^{\beta}
  & = & 1 + a_0 \exp( \beta \log s ) \cr
  & = & 1 + a_0 + a_0 \beta \log s + {\it O}( | \beta \log s |^2 ) .
\end{eqnarray}
Using $a_0 = \beta -1$ and $|\log s| \gg 1$, the $a_0 \beta \log s$
term is found to be dominant and we obtain
\begin{equation}
  <x^q> \simeq \frac{c_q}{ | a_0 \beta \log s | } .
\end{equation}
Thus \mbox{$<x^q>$} diverges logarithmically as $1 / | \log s |$.

\paragraph*{(b) $|\beta| \simeq q$}

Consider a parameter region where $\beta < 0$ and $|\beta| \simeq q$.
We further assume $| ( q - |\beta| ) \log s | \ll 1$.
\mbox{$<x^q>$} is then given by
\begin{equation}
  <x^q> \simeq \frac{c_q}{a_0} \left( s^{|\beta|} + a_q s^q \right)
  = \frac{c_q}{a_0} s^{|\beta|} \left( 1 + a_q s^{q - |\beta|} \right) .
\end{equation}
Expanding the right-hand side, we obtain
\begin{eqnarray}
  1 + a_q s^{q-|\beta|} & = & 1 + a_q \exp\left( ( q-|\beta| ) \log s \right) \cr
  & = & 1 + a_q + a_q ( q-|\beta| ) \log s \cr
  && \ \ \ \ \ \ \ \ \ + {\it O}( | ( q-|\beta| ) \log s |^2 ) .
\end{eqnarray}
By using $a_q = \frac{\beta+q}{1+q}-1 $ and $|\log s| \gg 1$, the
$a_q ( q-|\beta| ) \log s$ term is found to be dominant, and we obtain
\begin{equation}
  <x^q> \simeq \frac{c_q a_q}{a_0} | ( q-|\beta| ) s^{|\beta|} \log s | .
\end{equation}
Thus \mbox{$<x^q>$} diverges as $| s^{|\beta|} \log s |$.

\section{Exact treatment}

Next, we treat the effect of the additive noise without approximation.
The argument below is in parallel with the previous one, and their
results agree qualitatively, giving the same values of the exponents.
How to calculate the PDF follows the argument by Venkataramani
{\it et al.}~\cite{Venkataramani}.

\subsection{Fokker-Planck equation}

The Fokker-Planck equation corresponding to the Langevin equation
(\ref{eq:Langevin}) is given by
\begin{equation}
  \label{eq:FokkerPlanck}
  \frac{\partial}{\partial t} P(x, t) =
  - \frac{\partial}{\partial x} j(x, t) ,
\end{equation}
where $P(x, t)$ is the PDF of $x(t)$, and the flux $j(x, t)$ takes the
form
\begin{equation}
  j(x, t) = (\lambda_0 + D_{\lambda}) x P(x, t) - \frac{\partial}{\partial x} 
  [ ( D_{\lambda} x^2 + D_{\eta} ) P(x, t) ] .
\end{equation}
Reflective walls at $x = \pm 1$ are equivalent to imposing no-flux
boundary conditions at $x = \pm 1$, i.e.,
\begin{equation}
  j(x = \pm 1, t) = 0 .
\end{equation}

\subsection{Stationary PDF with a power-law tail}

We calculate the stationary solution $P(x)$ of the Fokker-Planck
equation (\ref{eq:FokkerPlanck}).
Stationarity condition ${\partial P(x, t)}/{\partial t} = 0$ gives
${\partial j(x, t)}/{\partial x} = 0$, i.e., $j(x) \equiv const.$, and
no-flux boundary conditions $j(x = \pm 1) = 0$ give $j(x) \equiv 0$.
Therefore, $P(x)$ obeys
\begin{equation}
  (\lambda_0 + D_{\lambda}) x P(x) - \frac{\partial}{\partial x} 
  [ ( D_{\lambda} x^2 + D_{\eta} ) P(x) ] = 0 .
\end{equation}
Solving this, we obtain
\begin{equation}
  P(x) = C (D_{\lambda} x^2 + D_{\eta})^{ \frac{\lambda_0}{2D_{\lambda}} - \frac{1}{2} } 
\end{equation}
as the stationary PDF, where $C$ is a normalization constant to be
determined from
\begin{equation}
  \int_{-1}^{1} P(x) dx = 2 \int_{0}^{1} P(x) dx = 1 .
\end{equation}
If we use the integral formula
\begin{equation}
  \displaystyle{ \int_0^1 (1+cx^2)^a dx = {_2F_1}(-a, \frac{1}{2}, \frac{3}{2}; -c) } ,
\end{equation}
where ${_2F_1}(a, b, c; z)$ is the hypergeometric function,
$C$ can be expressed as
\begin{equation}
  C = \left[ 2 D_{\eta}^{ \frac{\lambda_0}{2 D_{\lambda}} - \frac{1}{2} } \ 
    {_2F_1}( - \frac{\lambda_0}{2 D_{\lambda}} + \frac{1}{2}, \frac{1}{2}, \frac{3}{2};
    -\frac{D_{\lambda}}{D_{\eta}} ) \right]^{-1} .
\end{equation}
We define $\beta$ as the ratio of the average expansion rate
$\lambda_0$ to its fluctuation $D_{\lambda}$ as in the previous
calculation, and $\alpha$ as the exponent of $P(x)$, i.e.,
\begin{equation}
  \beta = \frac{\lambda_0}{D_\lambda} , \ \ \  
  \alpha = \frac{\beta - 1}{2} = \frac{\lambda_0}{2 D_{\lambda}} - \frac{1}{2} .
\end{equation}
Further, we define $s$ as the ratio of the strength $\sqrt{D_{\eta}}$
of the additive noise to the strength $\sqrt{D_{\lambda}}$ of the
multiplicative noise:
\begin{equation}
  s = \sqrt{\frac{D_{\eta}}{D_{\lambda}}} .
\end{equation}
Finally, the stationary PDF is expressed as
\begin{equation}
  \label{eq:PDF2}
  P(x) = \left\{
    \begin{array}{cc}
      \displaystyle{
        \frac{ (1 + \frac{x^2}{s^2})^{\alpha} }
        {2 {_2F_1}( -\alpha, \frac{1}{2}, \frac{3}{2}; -\frac{1}{s^2} )}
        } & (|x| < 1) \\
      \\
      0 & (|x| > 1)
    \end{array}
  \right. .
\end{equation}

This stationary PDF approaches a constant as $x \rightarrow 0$, and
when $s^2 \ll 1$, namely, when the additive noise is sufficiently
weaker than the multiplicative noise, the PDF approaches a power law
as $x \rightarrow \pm 1$:
\begin{equation}
  P(x) \sim \left\{
    \begin{array}{ll}
      const. & (x \rightarrow 0) \\
      \\
      |x|^{2 \alpha} & (x \rightarrow \pm 1) 
    \end{array}
  \right. .
\end{equation}
The exponent of this power law is given by
\begin{equation}
  2 \alpha = \beta - 1 = \frac{\lambda_0}{D_{\lambda}} - 1 .
\end{equation}
Thus, we obtain qualitatively the same PDF as the previous approximate
result (\ref{eq:PDF1}), in particular, a power law with the same
exponent. The crossover point between the constant region and the
power-law region of the PDF is found from the balance of the two terms
in the numerator of $P(x)$:
\begin{equation}
  1 \simeq \frac{x^2}{s^2} .
\end{equation}
The crossover thus occurs near
\begin{equation}
  s = \sqrt{\frac{D_{\eta}}{D_{\lambda}}} ,
\end{equation}
which is exactly the point at which we divided the domain of $x$ in
the previous approximate treatment.

\begin{figure}[htbp]
  \begin{center}
    \leavevmode
    \epsfxsize=8cm
    \epsffile{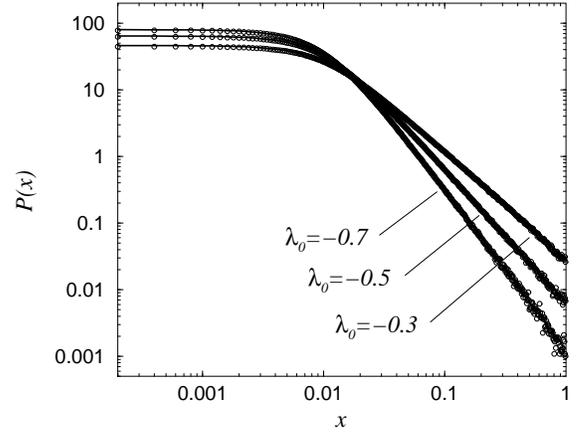}
  \end{center}
  \caption{PDFs $P(x)$ vs. $x$ obtained numerically($\circ$) and theoretically(---)
    for different values of $\lambda_0$. $D_\lambda = 0.5$ and $D_\eta
    = 0.00005$ are fixed.}
  \label{fig:3}
\end{figure}

\begin{figure}[htbp]
  \begin{center}
    \leavevmode
    \epsfxsize=8cm
    \epsffile{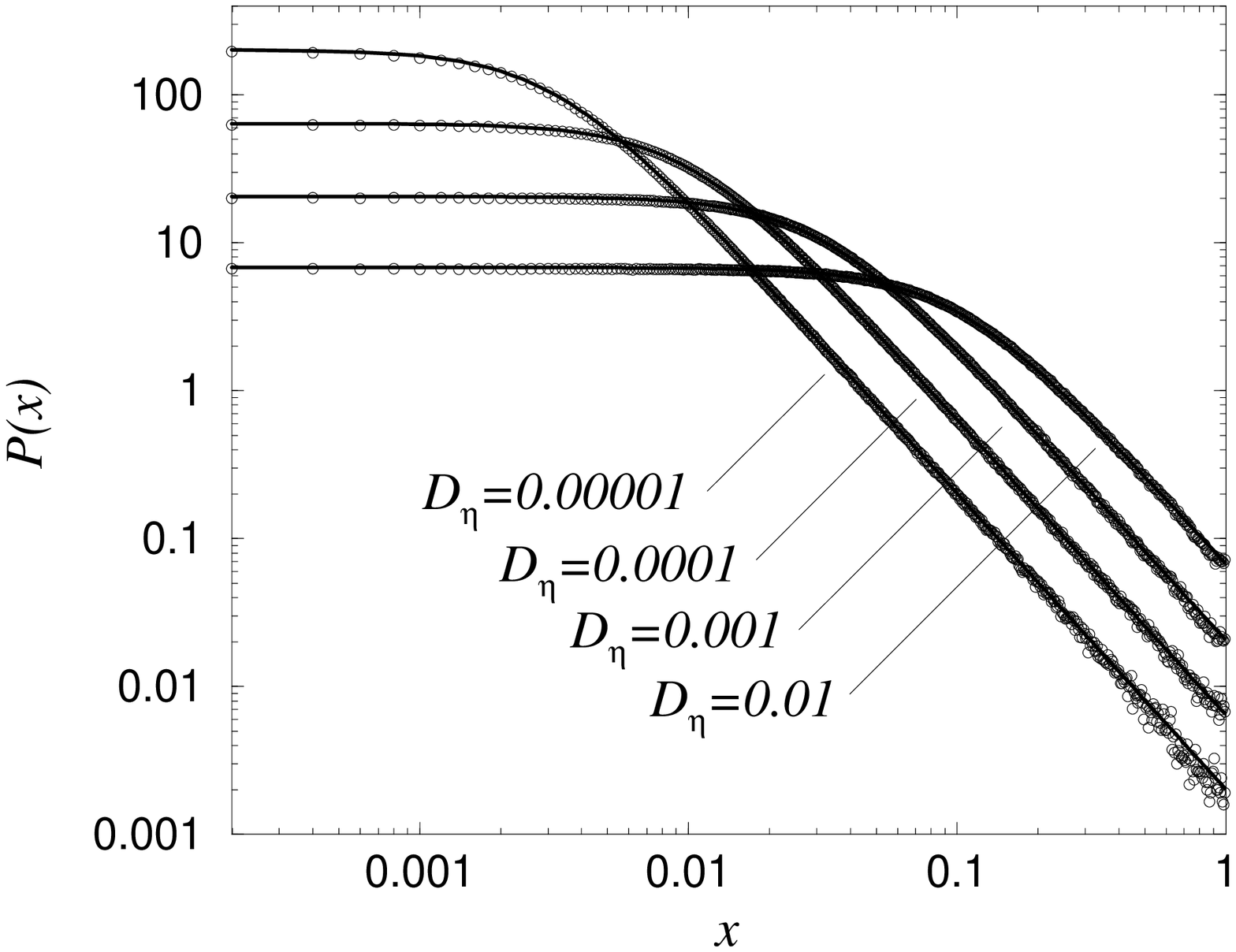}
  \end{center}
  \caption{PDFs $P(x)$ vs. $x$ obtained numerically($\circ$) and theoretically(---)
    for different values of $D_\eta$. $\lambda_0 = -0.5$ and
    $D_\lambda = 0.5$ are fixed.}
  \label{fig:4}
\end{figure}

We show the exact PDF (\ref{eq:PDF2}) in Fig.~\ref{fig:2}. The
approximate PDF (\ref{eq:PDF1}) reproduces the main features of the
exact one well.
In Fig.~\ref{fig:3} and \ref{fig:4}, we display two graphs of PDFs,
one obtained theoretically in (\ref{eq:PDF2}) and the other obtained
numerically by a direct simulation of the Langevin equation
(\ref{eq:Langevin}). Figure \ref{fig:3} illustrates PDFs for
different values of $\lambda_0$ with fixed $D_\eta$, while those in
Fig.~\ref{fig:4} are for different values of $D_\eta$ with fixed
$\lambda_0$. Each PDF takes a constant value near the origin, whereas
it obeys a power law otherwise. Their exponents vary with
$\lambda_0$, and the crossover position moves to the right with the
increase of the strength $\sqrt{D_\eta}$ of the additive noise.

\subsection{Moments}

\begin{figure}[htbp]
  \begin{center}
    \leavevmode
    \epsfxsize=8cm
    \epsffile{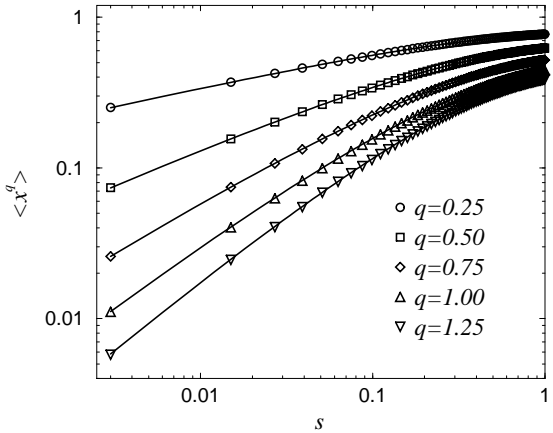}
  \end{center}
  \caption{Moments \mbox{$<x^q>$} vs. the strength $s$ of the additive noise
    obtained numerically(symbols) and theoretically(lines) for several
    values of $q$. The parameters are $\lambda_0 = -0.5$ and
    $D_\lambda = 0.5$.}
  \label{fig:5}
\end{figure}

The $q$-th moment \mbox{$<x^q>$} with respect to the stationary PDF
(\ref{eq:PDF2}) is calculated as
\begin{eqnarray}
  \label{eq:Mom_exact}
  <x^q> & = & \int_{-1}^{1} |x|^q P(x) dx = 2 \int_{0}^{1} x^q P(x) dx \cr
  & = & \frac{1}{1+q} \frac{_2F_1(-\alpha, \frac{1+q}{2}, \frac{3+q}{2}; -\frac{1}{s^2})}
  {{_2F_1}(-\alpha, \frac{1}{2}, \frac{3}{2}; -\frac{1}{s^2})} ,
\end{eqnarray}
where we used the integral formula
\begin{equation}
  \displaystyle{ \int_0^1 x^q (1+cx^2)^a dx
    = \frac{1}{1+q} {_2F_1}(-a, \frac{1+q}{2}, \frac{3+q}{2}; -c) } .
\end{equation}
In Fig.~\ref{fig:5}, we show the moments \mbox{$<x^q>$} obtained
theoretically in (\ref{eq:Mom_exact}) and compare them with those
obtained numerically.

\subsection{Asymptotic forms of the moments}

We investigate the asymptotic forms of the moment \mbox{$<x^q>$} in the limit
of small additive noise, $s \rightarrow 0$. As before, we consider
only the case $q>0$.

Using the asymptotic form of the hypergeometric function, i.e.,
\begin{eqnarray}
  {_2F_1}(a, b, c; z) \simeq \Gamma_1(a, b, c)(-z)^{-a} 
  + \Gamma_2(a, b, c)(-z)^{-b} \cr
  \cr
  (z \rightarrow \infty) ,
\end{eqnarray}
we can write the asymptotic form of \mbox{$<x^q>$} as
\begin{equation}
  <x^q> \simeq \frac{1}{1+q} \ 
  \frac{\Gamma_1(-\alpha, \frac{1+q}{2}, \frac{3+q}{2})}
  {\Gamma_1(-\alpha, \frac{1}{2}, \frac{3}{2})} \ 
  \frac{1+\Gamma_3(-\alpha, \frac{1+q}{2}, \frac{3+q}{2}) \ s^{\beta+q}}
  {1+\Gamma_3(-\alpha, \frac{1}{2}, \frac{3}{2}) \ s^{\beta}} ,
\end{equation}
where $\Gamma_1$, $\Gamma_2$, and $\Gamma_3$ are defined in terms of the
gamma function $\Gamma(a)$ as
\begin{equation}
  \Gamma_1(a, b, c) = \frac{ \Gamma(c) \Gamma(b-a) }{ \Gamma(b) \Gamma(c-a) } , \ 
  \Gamma_2(a, b, c) = \frac{ \Gamma(c) \Gamma(a-b) }{ \Gamma(a) \Gamma(c-b) } ,
\end{equation}
and
\begin{equation}
  \Gamma_3(a, b, c) = \frac{ \Gamma_2(a, b, c) }{ \Gamma_1(a, b, c) } .
\end{equation}

\begin{figure}[htbp]
  \begin{center}
    \leavevmode
    \epsfxsize=8cm
    \epsffile{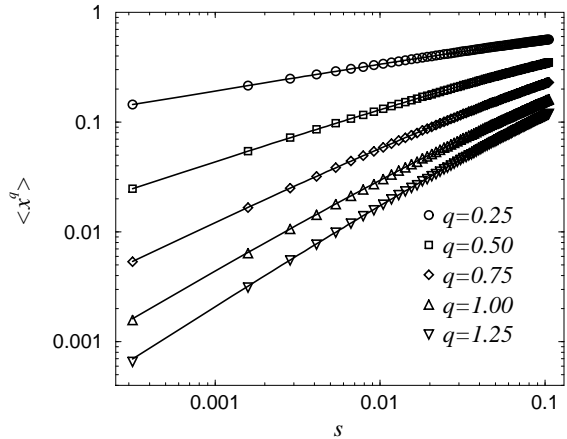}
  \end{center}
  \caption{Moments \mbox{$<x^q>$} vs. the strength $s$ of the additive noise
    obtained numerically(symbols) and theoretically(lines); a blowup
    of Fig. 5 for a small $s$ region.}
  \label{fig:6}
\end{figure}

Notice that here again we obtain the form already obtained in the
previous approximate calculation:
\begin{equation}
  <x^q> = c_q \frac{1 + a_q s^{\beta+q}}{1 + a_0 s^{\beta}} .
\end{equation}
However, the expressions for $c_q$ and $a_q$ are different, and now
given by
\begin{equation}
  c_q = \frac{1}{1+q} \frac{\Gamma_1(-\alpha, \frac{1+q}{2}, \frac{3+q}{2})}
  {\Gamma_1(-\alpha, \frac{1}{2}, \frac{3}{2})} , \ 
  a_q = \Gamma_3(-\alpha, \frac{1+q}{2}, \frac{3+q}{2}) .
\end{equation}
Using this form, and from exactly the same reasoning as before, we can
show that \mbox{$<x^q>$} asymptotically obeys a power law as
$s \rightarrow 0$:
\begin{equation}
  \label{eq:Mom_pow}
  <x^q> \simeq G_0 + G_1 s^{H(q)} .
\end{equation}
In Fig.~\ref{fig:6}, we show the moments \mbox{$<x^q>$} for small $s$
obtained both theoretically and numerically. Each moment shows
power-law dependence on the strength $s$ of the additive noise.

\subsection{Exponents and other asymptotic regimes}

The only difference between the above exact result and the previous
approximate result is in some coefficients involved. Since the
exponents are unchanged, the behavior of $H(q)$ is exactly the same as
the previous result. There is also no difference in that there exist
other asymptotic regimes near $\beta = 0$ or $|\beta| = q$, which is
clear if we notice $a_q = -1 + {\it O}(\beta)$ and $|\log s| \gg 1$.

\begin{figure}[htbp]
  \begin{center}
    \leavevmode
    \epsfxsize=8cm
    \epsffile{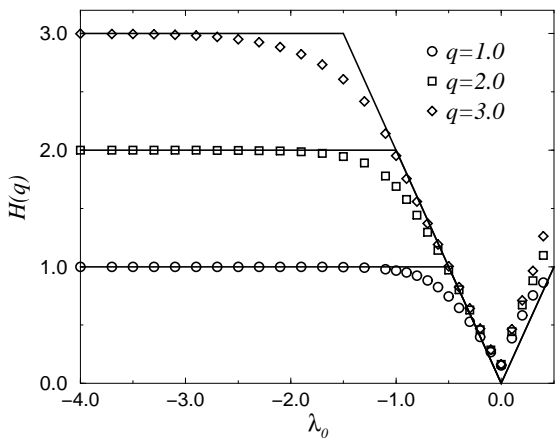}
  \end{center}
  \caption{Exponents $H(q)$ vs. $\lambda_0$ obtained numerically(symbols) and
    theoretically(lines) for $q = 1, 2, 3$. The value of $D_\lambda$
    is fixed to $0.5$.}
  \label{fig:7}
\end{figure}

In Fig.~\ref{fig:7}, the exponent $H(q)$ versus $\lambda_0$ obtained
theoretically in (\ref{eq:Exp_neg}) and (\ref{eq:Exp_pos}) are given
in comparison with those obtained numerically. Each $H(q)-\lambda_0$
curve is composed of two parts, i.e., a part where $H(q)$ varies in
proportion to $|\lambda_0|$ and that where $H(q)$ saturates to a
constant.
We estimated the exponents numerically by assuming a power law even in
the above-mentioned non-power-law asymptotic regimes. Therefore, the
estimated values there are naturally different from those expected
theoretically.

\section{Robustness of the power law}

Although we have treated only the Langevin equation
(\ref{eq:Langevin}) up to this point, the type of power law discussed
above appears in many other models. It is insensitive to the details
of the model such as the boundary conditions imposed, discreteness or
continuity of time, and the nature of noise terms. We thus discuss the
robustness of the power law here.

\subsection{Boundary conditions}

We treated the effect of the additive noise explicitly both in the
approximate and exact calculations. The crucial role of the additive
noise in generating a power-law PDF is to save the stochastic variable
from decaying completely by generating small fluctuations around the
zero value where a normal diffusion process dominates. This is the
reason why the usual approximation of replacing the additive noise
with an explicit lower bound of the variable works well.

Although we assumed that the upper bound of the stochastic variable is
simply given by the reflective walls, the result would not change
essentially if we replace it with some nonlinearity as given by a
$-x^3$ term, at least for not too large $q$. This is because the
dominant contribution to the $s$-dependence of \mbox{$<x^q>$} comes
from the region of large $P(x)$, i.e., that of small $x$ and not from
the large $x$ region near the upper bound.

\subsection{Discrete models}

Power laws also appear in discrete-time models, and their origin is
exactly the same as before. For example, in \cite{Kuramoto2} we
introduced a discrete time stochastic process
\begin{equation}
  x_{n+1} = e^{\lambda_{n}} x_n + O(x_{n}^2) + \eta_{n} ,
\end{equation}
where $n$ is the time step, $\lambda_{n}$ and $\eta_{n}$ are
noise. We approximated the additive noise term and the nonlinear term
as lower and upper reflective walls, and obtained a power-law
PDF. Furthermore, we obtained a power-law dependence of the moments of
$x$ on the position of the lower bound, i.e., the strength of the
additive noise.

\subsection{Nature of noise}

Of course, assuming a Gaussian-white noise is sometimes inadequate for
models of real systems. The power law of moments can also be seen in
some models with colored noises. For example, in \cite{Kuramoto1} we
introduced a stochastic process driven by a colored dichotomous noise
with lower and upper reflective walls. We obtained the power law of
moments with respect to the position of the lower wall as well.

\subsection{Numerical examples}

In order to demonstrate the robustness of the power law with regard to
the nature of noise terms and boundary conditions, we give a few
numerical examples below.

We numerically study a stochastic process
\begin{equation}
  \frac{d x(t)}{dt} = [ \lambda_0 + \lambda(t) ] x(t) + \eta(t) ,
\end{equation}
where $\lambda(t)$ is some colored noise, while $\eta(t)$ is a
Gaussian-white noise as before. We use three different types of noise
for $\lambda(t)$: (i) Gauss-Markov noise produced by an
Ornstein-Uhlenbeck process, (ii) dichotomous noise which takes values
$1$ or $-1$ with equal transition probabilities, and (iii) chaotic
noise produced by the Lorenz model. We normalize the average and
variance of each noise to $0$ and $1$ respectively, and use it for
$\lambda(t)$. We adjust $\lambda_0$ to a value where $x(t)$ shows
intermittency similar to Fig.~\ref{fig:1}.

\begin{figure}[htbp]
  \begin{center}
    \leavevmode
    \epsfxsize=8cm
    \epsffile{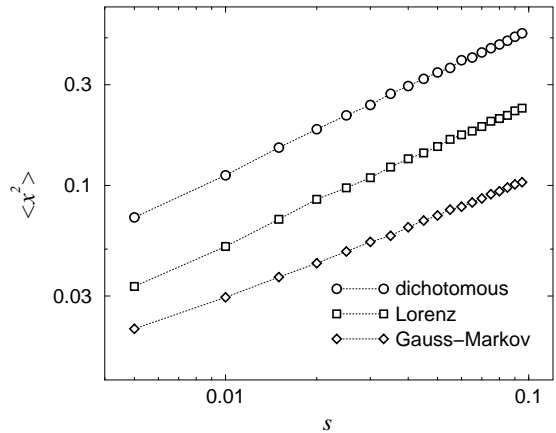}
  \end{center}
  \caption{Second moments \mbox{$<x^2>$} vs. the strength $s$ of the additive noise
    calculated for dichotomous noise, Gauss-Markov noise, and Lorenz
    noise. The value of $\lambda_0$ is $-0.5$ for the dichotomous and
    Gauss-Markov noise, and $-0.3$ for the Lorenz noise. Each line is
    shifted upwards or downwards to avoid overlap.}
  \label{fig:8}
\end{figure}

In Fig.~\ref{fig:8}, we show second moments \mbox{$<x^2>$} versus the
strength $s$ of the additive noise obtained for each type of noise.
We can see a power-law dependence of the moments \mbox{$<x^2>$} on the
strength $s$ of the additive noise in the small $s$ region.
We also studied the case where the saturation of $x$ is not due to the
reflective walls but a nonlinearity $-x^3$, and found that a power law
with the same exponent still holds.
We also confirmed that the PDF has a power-law tail for each type of
noise.

Since the noise is colored, it would be difficult to predict the
values of the exponents of the moments from the previous theory. In
order to achieve this, some sort of renormalization procedure, like
the one done in \cite{Pikovsky}, must be invoked to give effective
$\lambda_0$ and $D_\lambda$. It is beyond the scope of this paper.

\section{Some related systems}

\subsection{Noisy on-off intermittency}

Since noisy on-off intermittency is a typical phenomenon with the
mechanism of generating the power law of moments, we briefly discuss
it here.

On-off intermittency is observed where a chaotic attractor becomes
marginally stable with respect to disturbances transversal to the
invariant manifold in which the chaotic attractor is embedded. This
type of instability is called a blowout bifurcation.
The system then alternates between two phases intermittently: a
laminar phase where the system stays practically on the invariant
manifold, and a burst phase where the deviation from the invariant
manifold grows suddenly.
The mechanism responsible for the on-off intermittency is that the
distance between the orbit and the invariant manifold is governed by a
multiplicative process with a chaotically changing multiplier.
Therefore, the corresponding suitable mathematical model is a RMP,
where the fast chaotic motion of the multiplier is considered as a
multiplicative noise.

Platt {\it et al.}~\cite{Platt} investigated a situation where weak
additive noise is also present in the system. They found that the
intermittency, which was originally observed only in a narrow
supercritical-side region of the blowout bifurcation, can be observed
in a wider region including the subcritical side of the blowout
bifurcation. This is called noisy on-off intermittency. In order to
explain this phenomenon, they proposed a RMP model which incorporates
the effect of the additive noise approximately by introducing a lower
bound to the conventional model of on-off intermittency.
Venkataramani {\it et al.}~\cite{Venkataramani} and \v{C}enys
{\it et al.}~\cite{Cenys} also analyzed similar models and explained some
features of the noisy on-off intermittency.
However, the power law of moments which we described above has not
fully been investigated. Thus, we give an example below.

\subsection{Coupled chaotic elements}

Two identical chaotic elements coupled with each other~\cite{Fujisaka}
is a typical system which shows on-off intermittency.
A frequently used model is
\begin{equation}
  \left\{
    \begin{array}{rcl}
      {\dot {\bi X}_1} &=& {\bi F}({\bi X}_1) + k ( {\bi X}_2 - {\bi X}_1 ) \\
      \\
      {\dot {\bi X}_2} &=& {\bi F}({\bi X}_2) + k ( {\bi X}_1 - {\bi X}_2 )
    \end{array}
  \right. ,
\end{equation}
where ${\dot {\bi X}} = {\bi F}({\bi X})$ gives the dynamics of the
individual element, and $k$ the coupling strength.
These two elements synchronize with each other when $k$ is larger than
a certain critical coupling strength $k_c$.
The difference ${\bi x} = {\bi X}_2 - {\bi X}_1$ is driven
multiplicatively by the chaotic motion of the elements as
\begin{equation}
  {\dot {\bi x}} = \left[ D{\bi F}(\frac{{\bi X}_{1}+{\bi X}_{2}}{2})
    - 2 k{\bi I} \right] \cdot {\bi x} + O({\bi x}^2) ,
\end{equation}
where $D$ means differentiation and ${\bi I}$ the unit matrix.
Slightly below $k_c$, ${\bi x}$ exhibits on-off intermittency. If we
further apply some weak additive noise, ${\bi x}$ exhibits
intermittent behavior above $k_c$ as well.
If we vary the strength of the additive noise and measure the $q$-th
moment \mbox{$<x^q>$} of a certain component of ${\bi x}$, it is
expected to behave like \mbox{$<x^q>$}$ \simeq G_0 + G_1 s^{H(q)}$ as
far as $s$ is sufficiently small.

\begin{figure}[htbp]
  \begin{center}
    \leavevmode
    \epsfxsize=8cm
    \epsffile{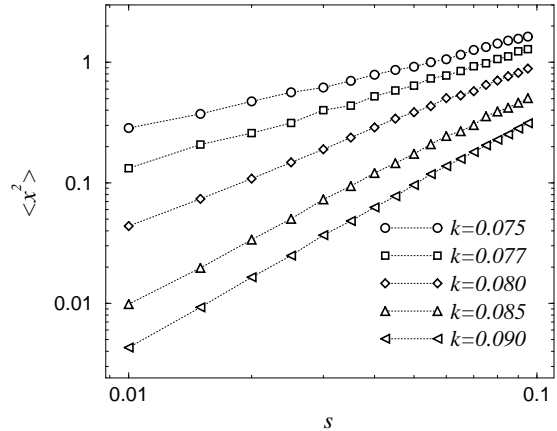}
  \end{center}
  \caption{Second moments \mbox{$<x^2>$} vs. the strength $s$ of the additive noise
    of coupled R\"ossler oscillators obtained for several values of $k$.}
  \label{fig:9}
\end{figure}

As an example, we numerically calculate a pair of R\"ossler
oscillators coupled with each other, and with weak additive noise
applied only to the first component of the state variables:
\begin{equation}
  \left\{
    \begin{array}{rcl}
      \dot{x}_{1}(t) &=& -y_{1}+z_{1} + k (x_{2} - x_{1}) + s \xi_{1}(t) \\
      \\
      \dot{y}_{1}(t) &=& x_{1}+0.3y_{1} + k (y_{2} - y_{1}) \\
      \\
      \dot{z}_{1}(t) &=& 0.2+x_{1}z_{1}-5.7z_{1} + k (z_{2} - z_{1})
    \end{array}
  \right.
  ,
\end{equation}
and
\begin{equation}
  ( 1 \longleftrightarrow 2) ,
\end{equation}
where $\xi_{1, 2}(t)$ are Gaussian-white noise of average $0$ and
variance $1$, and $s$ controls their strength.
We set the coupling strength $k$ to the value slightly above $k_c$,
where the system shows noisy on-off intermittency. Figure \ref{fig:9}
shows second moments \mbox{$<x^2>$} of the difference between the
first components $x = x_{2} - x_{1}$ versus the noise strength $s$,
obtained for some values of the coupling strength $k$. As expected,
they show power-law dependence on $s$, and their exponents vary with
$k$.

We also observed the power law of moments in a system of coupled
chaotic elements with a slight parameter mismatch~\cite{Yamada} in
place of a weak additive noise, where the parameter mismatch plays a
role similar to the additive noise. The power law of moments is also
observed in systems of coupled maps~\cite{Mishiro}.

\subsection{Spatially distributed chaotic elements}

The power-law spatial correlation which we studied in the previous
papers~\cite{Kuramoto1,Kuramoto2} turns out to be the same as the
power law described above, with a proper physical interpretation of
some variables involved.
The systems we treated previously are the populations of spatially
distributed chaotic elements. The elements are driven by a field
produced by non-local coupling, which is spatially long-waved and
temporally irregular.

If we consider the difference $x(r) = X(r) - X(0)$ between the
amplitude $X$ of two elements which are separated by a short distance
$r$, its moments $<x(r)^{q}>$ are directly related to the spatial
correlations of $X$. They are similar to the structure functions in
the context of fluid turbulence or the height-height correlations in
the context of fractal surface growth.
The dynamics of $x(r)$ is given by a RMP with a weak additive noise,
where the multiplier is the local Lyapunov exponent fluctuating
randomly due to the chaotic motion of the elements, and the weak
additive noise comes from the small difference in the strength of the
applied field between the two points under consideration.

Since the strength $s$ of the additive noise should be of the order of
$r$ due to the assumed smooth spatial variation of the applied field,
the power law of moments as a function of $s$ is now interpreted as a
power law of moments of the amplitude difference $x$ as a function of
the mutual distance $r$, i.e.,
\begin{equation}
  <x(r)^q> \sim G_0 + G_1 s^{H(q)} \sim G_0 + G_2 r^{H(q)} ,
\end{equation}
where $G_0$, $G_1$, and $G_2$ are constants.

When the average Lyapunov exponent is negative, $G_0$ vanishes and the
exponent $H(q)$ is given by Eq. (\ref{eq:Exp_neg}). It shows a
``bifractality'' similar to that known for Burgers equation, which
implies an underlying intermittent structure.

\section{Conclusion}

As in the models for noisy on-off intermittency and economic activity,
a stochastic process driven by multiplicative and weak additive noise
shows a power-law PDF.
The PDF consists of a constant part and a power-law part,
and their boundary moves with the strength $s$ of
the additive noise in such a way that its distance from the origin
is proportional to $s$.
This systematic dependence on $s$ causes the power law of the moments
with respect to the strength of the additive noise.

In order to study this phenomenon in further detail, we introduced a
Langevin equation (\ref{eq:Langevin}) with multiplicative and additive
noise terms as a general model for a stochastic process of this type.
We analyzed its stationary state theoretically and numerically,
and found that this model actually reproduces the power law of moments.
Furthermore, by comparing the approximate and exact treatments of the
effect of the additive noise, the usual approximation of the additive
noise by introducing a lower bound of the amplitude was justified.

Although we restricted our study to the Langevin equation
(\ref{eq:Langevin}) for the sake of precise argument, the power law
itself is not sensitive to the details of the model employed, and can
appear robustly in many stochastic processes driven by multiplicative
and weak additive noise.
We demonstrated such robustness numerically for some different models,
where the system is driven by noises which are not Gaussian-white.
Furthermore, as some typical realizations of this type of power law,
we discussed the power law of moments in noisy on-off intermittency,
and the power-law spatial correlation functions in the spatio-temporal
chaotic regime of non-locally coupled systems, which gives some more
insight into the power law of the spatial correlation function.

As we already noted, the power law of moments seems to be a general
phenomenon appearing in various systems over a wide range of
parameters. This mechanism of generating a power law seems to be quite
universal, and some of the power laws observed in the real world may
belong to this class.

\acknowledgments
The author is very grateful to Y. Kuramoto, P. Marcq, S. Kitsunezaki,
T. Mishiro, Y. Sakai, and the members of the Nonlinear Dynamics Group
of Kyoto University for valuable discussions and advice.

\end{document}